\documentclass[twocolumn,showpacs,preprintnumber,amsmath,amssymb,floatfix]{revtex4}

\usepackage{graphicx}
\usepackage{psfrag}
\usepackage{dcolumn}
\usepackage{bm}

\begin{document}

\title{Non-equilibrium Plasmons in a Quantum Wire Single Electron Transistor}

\author{Jaeuk U. Kim}
\author{Ilya V. Krive}
  \altaffiliation[Also at ]{
   B.I. Verkin Institute for Low Temperature Physics and Engineering, 
   Lenin ave., 47, Kharkov 61103, Ukraine}
\author{Jari M. Kinaret}
\affiliation{
Department of Applied Physics, Chalmers University of Technology 
and G\"{o}teborg University, SE-412 96 Gothenburg, Sweden}

\date{\today}

\begin{abstract}

We analyze a single electron transistor composed of two semi-infinite one
dimensional quantum wires  and a relatively short segment between them.
We describe each wire section by a Luttinger model, and treat 
tunneling events in the sequential approximation when the system's
dynamics can be described by a master equation. We show that
the steady state occupation probabilities in the strongly interacting regime 
depend only on the energies of the states and follow a universal form that depends on
the source-drain voltage and the interaction strength. 

\end{abstract}


\maketitle


Strongly interacting one dimensional (1D) electron systems 
that can be described by Tomonaga-Luttinger (TL) model \cite{tomonaga50} are 
under great interest due to recent experimental observations especially 
in the systems of single-walled carbon nanotubes (SWNT)
\cite{tans97,egger97,kane97,bockrath99}.
The Luttinger Liquid (LL), which is generalization of the TL model, 
is characterized by the absence of fermionic quasiparticles; instead, 
the elementary excitations take the form
of bosonic charge and spin density waves \cite{voit95}.

\begin{figure}[htbp]
\begin{center}
\includegraphics[width=6cm,height=2.5cm]{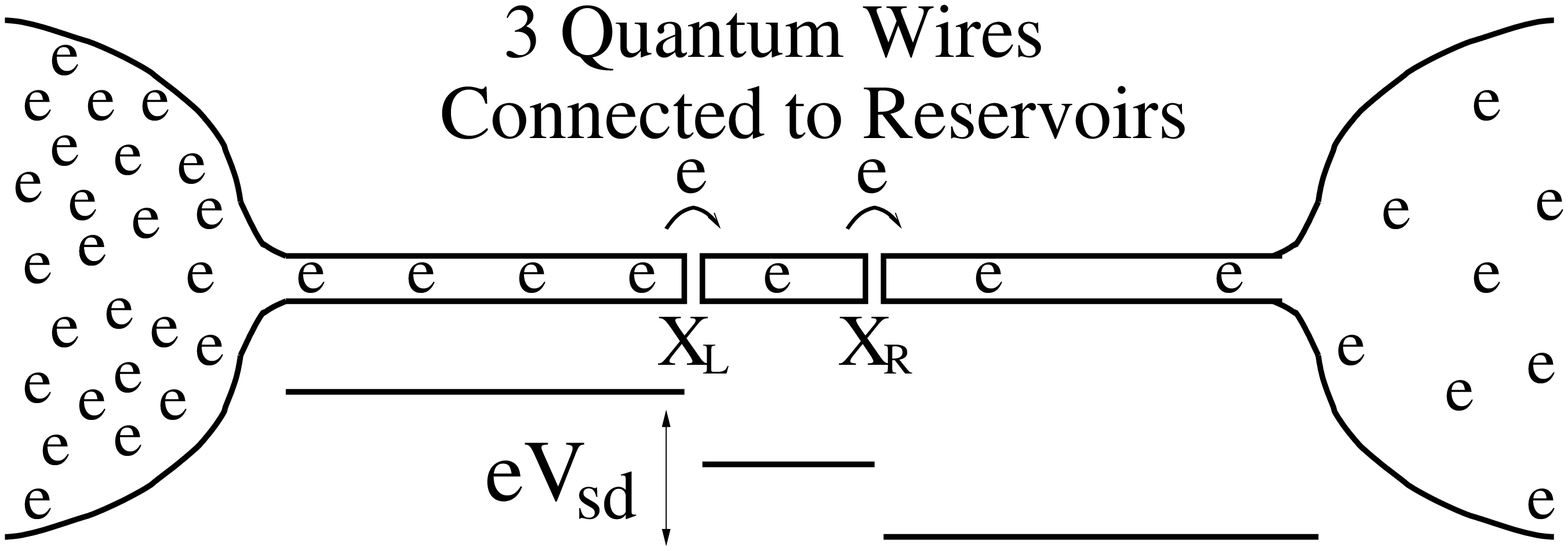}
\caption{Model system.
Two long wires are adiabatically connected to reservoirs and a short wire is 
connected to the two by weak coupling. }
\label{fig:model}
\end{center}
\end{figure}

Motivated by recent experiment by Postma {\em et al.} \cite{postma01}, 
we theoretically investigate a 1D single electron transistor (SET). 
Schematic description of the system is that a finite wire segment, 
later called a quantum dot, is weakly coupled to  
two long wires as depicted in Fig. \ref{fig:model}. 
Two point-like impurities or other defects in a long nanotube 
may result in equivalent physics. We model the system as a finite LL segment 
and two semi-infinite LL leads.  The leads are connected to electron 
reservoirs which keep them in internal equilibria. The chemical potentials 
of the leads are controlled by source-drain voltage ($V_{sd}$), 
and we assume that tunneling barriers between leads and dot are high enough 
that sequential tunneling is the dominant charge transport mechanism. 
Closely related studies based on a different set of approximations were
recently published by Braggio and co-workers \cite{braggio00,braggio01}.
If the tunneling barriers are lower, or if sequential tunneling is blocked
(Coulomb blockade), correlations between tunneling events across
the left and right junctions become important. Transport in this regime
has been discussed  by Thorwart {\em et al.} \cite{thorwart02}.

In this Letter, we analyze the steady-state occupation probabilities of the 
many-body states on the dot.  We will show that 
in the strongly interacting regime they are characterized by a {\em universal} 
distribution that depends on the applied voltage and the interaction strength. 


Carbon nanotubes have four transport channels including spin degeneracy \cite{egger97,kane97}. 
The electron transport through quantum wires (QWs) is determined mostly by charge effects so 
it is reasonable to consider, as a first approximation \cite{bockrath99},
 the contribution of the total charge  sector to SET characteristics,
and postpone the discussion of the role of the other channels to a later date.
The effective Hamiltonian $\hat H=\hat H_{0}+\hat H_T$ for this  branch is
\begin{equation}
\begin{array}{l}
\hat{H}_{0} = \begin{displaystyle}\sum_{l=(L,D,R)}\frac{\pi\hbar v_F}{gL_l}
\sum_{m=1}^{\infty}m \hat b^{\dagger}_{l,m}\hat b_{l,m} 
+ \frac{\pi\hbar v_F}{2g^2 L_D}(\hat N-N_G)^2, \end{displaystyle}\\
\hat H_T = \begin{displaystyle} \sum_{\lambda=(L,R)}\sum_{a=(i,\sigma)}
\left[ t_\lambda \hat \psi^\dagger _{D,a} 
(X_\lambda)\hat \psi_{\lambda,a} (X_\lambda) +H.C. \right]. \end{displaystyle}
\end{array}
\label{eq:Hamiltonian}
\end{equation}
Here $\hat H_0$ is the bosonized LL Hamiltonian 
and the tunneling Hamiltonian $\hat H_T$ describes 
 {\em bare electrons} tunneling through barriers at the points $X_L$ and $X_R$,
 where indices $i$=1,2 and $\sigma$=$\pm$ identify bare electrons.
The parameter $g$  indicates interaction
strength, $g$=1 for non-interacting electrons
 and $g$$\rightarrow$$0$ for strong repulsive interaction.
The term $\hat E_C$=$\frac{\pi\hbar v_F}{2g^2 L_D}(\hat N-N_G)^2$ is on-dot zero-mode 
contribution including the Coulomb charging energy, where $N_G$ is 
a dimensionless gate voltage control parameter in the range $N_G$$\in$[0,1].
Leads are assumed to be semi-infinite with continuous spectra; 
the dot has finite size $L_D$ and discrete energy spectrum with
the plasmon level spacing $\Delta E = \pi\hbar v_F/gL_D$. 

We describe the system's dynamics using 
a master equation within a sequential tunneling regime where uncorrelated 
single electron hoppings are dominant transport mechanism \cite{furusaki98,komnik02},
\begin{equation}
\begin{array}{l}
(\partial/\partial t) P(N,\{n\}) \\
\begin{displaystyle}=\sum_{N^\prime = N\pm 1}\sum_{\{n^\prime\}}
\left[ P(N^\prime,\{n^\prime\}) \Gamma(N^\prime,\{n^\prime\} \rightarrow N,\{n\}) \right. 
\end{displaystyle}\\
\begin{displaystyle}
~~~~~~~~~~~~~~~~~\left. -P(N,\{n\})\Gamma(N,\{n\} \rightarrow N^\prime,\{n^\prime\}) \right],
\end{displaystyle}
\end{array}
\label{eq:Master}
\end{equation}
where the variables $(N,\{n\})$ represent a state of the dot with $N$ excess
electrons  and a set of the occupation numbers $\{n\}=(n_1,n_2,\cdots,n_m,\cdots)$
of the plasmon states with momenta $p_m=m\cdot\pi\hbar/L_D$, $P$ are the occupation 
probabilities of the dot-states $(N,\{n\})$, and
$\Gamma = \Gamma_L+\Gamma_R$ are the transition rates 
due to tunneling through junctions  at $X_L$ and $X_R$.
The lead degrees of freedom were integrated out since
the leads are locally in equilibria.

If the transition rates are known, 
it is straightforward to determine steady-state occupation probabilities 
by solving master equation (\ref{eq:Master}) for $(\partial/\partial t) P(N,\{n\}) = 0$ 
 with the constraint $\sum_{N,\{n\}} P(N,\{n\})=1$.

The transition rates $\Gamma_{L/R}$ are evaluated within the golden rule approximation 
with respect to the tunneling Hamiltonian $\hat H_T$ using the standard bosonization 
technique  \cite{voit95}.
As a function of the states on the dot and their energies the rates are given by
\begin{equation}
\begin{array}{l}
\Gamma_{L/R}(N,\{n\} \rightarrow N^\prime,\{n^\prime\}) ={8\pi|t_{L/R}|^2}/ {\hbar} \\
\times \gamma(E_D(N^\prime,\{n^\prime\})-E_D(N,\{n\})\mp (N^\prime- N) eV_{L/R})  \\
  \times |\langle N^\prime,\{n^\prime\}|
\psi^\dagger_D\delta_{N^\prime,N+1}+\psi_D\delta_{N^\prime,N-1}| N,\{n\}\rangle|^2,
\end{array}
\label{eq:Gamma}
\end{equation}
where ``$-/+$'' sign corresponds to index ``$L/R$'' 
and the voltage drops across junctions are $V_{L/R}=\frac{R_{L/R}}{(R_{L}+R_{R})}V_{sd}$ 
for the tunneling resistances $R_{L/R} \propto |t_{L/R}|^{-2}$ when dc source drain
voltage $V_{sd}$ is supplied.
Lead contributions are given by a spectral function $\gamma(\epsilon)$
\cite{furusaki98,braggio01}
\begin{equation}
\begin{array}{rl} 
\gamma(\epsilon)= & \frac{1}{2\pi\hbar}\int_{-\infty}^{\infty}
dt e^{i\epsilon t}\langle \Psi_{\lambda,a}(X_\lambda,0)
\Psi_{\lambda,a}^\dagger (X_\lambda,t)\rangle \\
=& \begin{displaystyle}
\frac{1}{2\pi\hbar}\frac{1}{\pi v_F}\left(\frac{2\pi \Lambda_g}{\hbar v_F\beta}
\right)^\alpha\left|\Gamma(\frac{\alpha+1}{2}+i\frac{\beta\epsilon}{2\pi})\right|^2
\frac{e^{-\beta\epsilon/2}}{\Gamma(\alpha+1)}  
\end{displaystyle}
\end{array}
\label{eq:gamma}
\end{equation}
Here $\alpha=(g^{-1}-1)/4$, appropriate for an end-contacted four-channel wire 
such as SWNT  \cite{kane97},  $\beta=1/k_BT$ is the
inverse temperature in the leads, and $\Lambda_g=\Lambda g^{1/(1-g)}$ where $\Lambda$
is a short wavelength cut-off.
At zero temperature, $\gamma(\epsilon)$ is proportional to a power of energy, 
$\lim_{T\rightarrow 0}\gamma(\epsilon) \propto \Theta(-\epsilon) |\epsilon|^\alpha$. 
The dot energy $E_D$ is obtained by solving the eigenvalue problem 
of the Hamiltonian (\ref{eq:Hamiltonian}), and is given by a sum 
of the zero mode energy $E_C(N)= (\pi\hbar v_F/2g^2L_D)\cdot(N-N_G)^2$ and 
the plasmonic excitation energy $E(\{n\})=(\pi\hbar v_F/gL_D)\cdot\sum_m m\cdot n_m$,
\begin{equation}
E_D(N,\{n\}) = \frac{\pi\hbar v_F}{L_D}\left[\frac{(N-N_G)^2}{2g^2}
+\sum_{m=1}^\infty \frac{m\cdot n_m}{g}\right]
\label{eq:E_D}
\end{equation}
The zero-mode overlap is unity for $N^\prime=N\pm 1$ and
vanishes otherwise, and the overlaps of the plasmonic states can be
written in terms of Laguerre polynomials as \cite{schwinger53}
\begin{equation}
\begin{array}{l}
|\langle \{n^\prime\}|\psi^\dagger_{D}|\{n\}\rangle|^2 
=\frac{1}{L_D}\left(\frac{\pi \Lambda}{L_D}\right)^{\alpha} \times\\
\begin{displaystyle}
 \prod_{m=1}^{\infty} \left(\frac{1}{4mg}\right)^{|n_m^\prime-n_m|} 
\frac{n_m^{(<)} !}{n_m^{(>)} !}
\left[ L^{|n_m^\prime-n_m|}_{n_m^{(<)}}\left(\frac{1}{4mg}\right)\right]^2
\end{displaystyle}
\end{array}
\label{eq:Laguerre}
\end{equation}
where $n_m^{(<)}=\min(n_m^\prime,n_m)$ and $n_m^{(>)} = \max(n_m^\prime,n_m)$. 
Notice that at low energies only the first few occupations $n_m$ and $n^\prime_m$
in the product in Eq. (\ref{eq:Laguerre}) differ from zero, yielding non-unity factors to 
the transition rate.

With the analytically calculated transition rates 
(Eqs. (\ref{eq:Gamma})-(\ref{eq:Laguerre})), we solve the master equation numerically 
and determine occupation probabilities of the many-body states.
Hereafter we  will present numerical results for the case with symmetric 
barriers,  the dot size of $L_D=50nm$ and the Fermi velocity of 
$v_F=8\times10^5m/s$ \cite{bockrath99}. 

\begin{figure}[htbp]
\begin{center}
\psfrag{xlab_A0}{0} \psfrag{xlab_A1}{0.1} \psfrag{xlab_A2}{0.2} \psfrag{xlab_A3}{0.3}
\psfrag{xlab_A4}{0.4}
\psfrag{xlab_AE}{E(\{n\}) [eV]}
\psfrag{ylab_A10}{10}
\psfrag{ylab_A2}{$^{\mbox{-}2}$}
\psfrag{ylab_A4}{$^{\mbox{-}4}$}
\psfrag{glab_A}{g=1}
\psfrag{xlab_B0}{0} \psfrag{xlab_B3}{3} \psfrag{xlab_B6}{6} \psfrag{xlab_B9}{9}
\psfrag{xlab_B12}{12}
\psfrag{xlab_BE}{E(n) [eV]}
\psfrag{ylab_B0}{0} \psfrag{ylab_B2}{.2} \psfrag{ylab_B4}{.4} \psfrag{ylab_B6}{.6}
\psfrag{ylab_B8}{.8}
\psfrag{glab_Bin12}{.12} \psfrag{glab_Bin24}{.24} \psfrag{glab_Bin40}{.40} \psfrag{glab_Bin56}{.56}
\psfrag{glab_Bin72}{.72} \psfrag{glab_Bin100}{g=1.00} \psfrag{glab_Bin88}{.88}
\includegraphics[width=7cm,height=10.5cm]{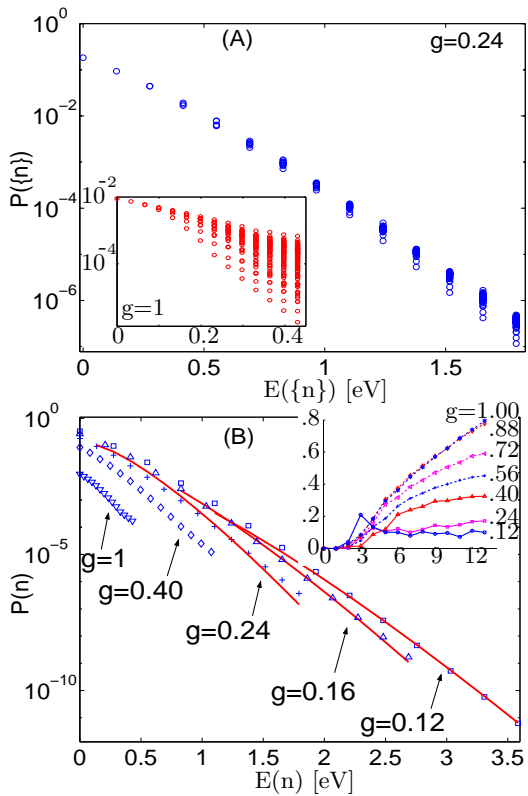}
\caption{(A) The occupation probabilities $P(\{n\})\equiv P(N$=$0,\{n\})$ 
as a function of the energies $E(N$=$0,\{n\})$ of the many-body states on the dot 
for $g=0.24$ (SWNT) and for $g=1$ (non-interacting) in the inset. 
The degeneracy of the energy state $E(n)$ is
$D(n) \simeq {\exp(\pi\sqrt{2n/3})}{/(4\sqrt{3}n)}$, where $n=\sum_m m\cdot n_m$ 
is the dimensionless energy
(circles ($\circ$) represent individual many-body states).
(B) Averaged probabilities $P(n)=\langle P(\{n\})\rangle_{n}$ as a function of
the energy levels $E(n)$. Solid lines are analytic approximations
$P(n)\propto n^{-\frac{3}{2}\frac{\alpha+1}{n_{sd}}n}$ 
where $n_{sd}=eV_{sd}\cdot(\pi\hbar v_F/gL_D)^{-1}$.
Inset shows the relative standard deviations of the probabilities,
$\langle\delta P(\{n\})\rangle_n/P(n)$, as a function of the dimensionless 
energy $n=E(n)\cdot(\pi\hbar v_F/gL_D)^{-1} $
 for $g$-values ranging from $g=1$ to $g=0.12$.
Implicit parameters are $T=50mK, V_{sd}=g^{-1}\times 0.22 V,$ and {$N_G$=0.5}.
 }
\label{fig:PnallvsEn}
\end{center}
\end{figure}

Fig. \ref{fig:PnallvsEn} shows (A) the occupation probabilities $P(\{n\})$ {\em vs.}
the energies $E(\{n\})$ of the many-body states of ($N$=0,$\{n\}$) and (B) 
the averaged probabilities $P(n)=\langle P(\{n\})\rangle_n$ of degenerate 
states with energies $E(n)$.  The degeneracy $D(n)$ of 
$E(n)=n\cdot \pi\hbar v_F/gL_D$ is the number of the many-body occupation 
configurations $\{n\}$
that satisfy $E(\{n\})=E(n)$, {\em i.e.}, 
$\sum_m m\cdot n_m = n$, which asymptotically follows the Hardy-Ramanujan formula 
\cite{ramanujan18}
\begin{equation}
D(n) \simeq e^{\pi\sqrt{2n/3}}/(4\sqrt{3}n). 
\label{eq:HR}
\end{equation}

As seen in Fig. \ref{fig:PnallvsEn}(A), 
there is a dramatic change of the occupation probability distribution between 
a non-interacting ($g$=1) and a strongly interacting ($g\lesssim 0.24$) quantum dot.
In the non-interacting case the occupation probabilities of degenerate many-body states 
vary widely as is natural since electrons tunnel independently of each other
(apart from the Pauli principle); typically states with many long
wavelength plasmons are more likely than states with 
few short wavelength excitations.  

In contrast, in the strongly interaction regime 
all degenerate states are almost equally likely (variation of about 10\% only,
see the inset of Fig. \ref{fig:PnallvsEn}(B)) and the occupation probability
decreases nearly exponentially with energy. The suppression of 
variations originates from interactions in the {\em central island} 
\cite{jaeuk02} and can be understood as a consequence of the separation
of zero-modes and bosonic excitations in the strongly interacting limit
even for a small system. 
The inclusion of spin effects in the central segment (spin is already included in the leads)
introduces non-interacting degrees of freedom that results in increased
fluctuations which can be suppressed by a magnetic field.
In  Fig. \ref{fig:PnallvsEn}(B), we show that the average occupation probabilities 
$P(n)=\langle P(\{n\})\rangle_{n}$ 
decay nearly exponentially
even in the weakly interacting regime, $g \gtrsim 0.4$,
where the probabilities of degenerate states fluctuate strongly.

The nearly exponential decay of the average probabilities 
suggests a 
 {\em Gibbs-like distribution}. However, there is no obvious reason why this
non-equilibrium distribution should follow an equilibrium form. 
We obtain an analytic approximation to $P(n)$ at $T$=$0$
by setting the on-dot transition elements in (\ref{eq:Laguerre}) to unity
and considering the scattering-in and scattering-out processes
for a particular many-body state $\{n\}$.
In this approximation, all degenerate states have equal probabilities and 
the total scattering rates at a specific energy are given by the  product
of the lead contribution (\ref{eq:gamma}), which follows
a power-law, and the degeneracy (\ref{eq:HR}).
The master equation now reads
\begin{equation}
\frac{\partial}{\partial t} P(n)=\sum_{n^\prime}
\bigl[ P(n^\prime) D(n^\prime)\Gamma(n^\prime \rightarrow n) 
-P(n)D(n^\prime)\Gamma(n \rightarrow n^\prime) \bigr]
\label{eq:Master2}
\end{equation}
where the total charge labels $N$ have been suppressed.
By solving the master equation using saddle point integral approximations,
we obtain a differential equation for $\ln[P(n)]$ and find that for a large
$E(n)$, $P(n)$ is 
\begin{equation}
P_{univ}(n) =Z^{-1}\cdot n^{-\frac{3}{2}\frac{\alpha+1}{n_{sd}}n}
=Z^{-1}\cdot e^{-\frac{3}{2}\frac{\alpha+1}{n_{sd}}n\log n}
\label{eq:P_analytic}
\end{equation}
where  $n_{sd}=eV_{sd}\cdot(\pi\hbar v_F/gL_D)^{-1}$ is the dimensionless bias voltage 
and $Z$ is the partition function.
This distribution is {\em universal} in the sense that it does not depend on
the precise form of the Hamiltonian of the dot as long as the excitation spectrum 
is bosonic and has linear dispersion.
Over limited energy ranges, it resembles 
a {\em Gibbs distribution} with the effective temperature
\begin{equation}
k_BT_{eff}(n) \approx -\left(\frac{\partial\ln P}{\partial E}\Big|_{E(n)} \right)^{-1}
=\frac{2}{3}\frac{eV_{sd}}{\alpha+1}\frac{1}{1+\log n}
\label{eq:kTeff}
\end{equation}
In the range of our interest $g \lesssim 0.24$ ($g\simeq$ 0.24 for a SWNT device 
\cite{bockrath99}) and $n\sim 10$, this is approximately 
\begin{equation}
k_BT_{eff}\simeq C(n)\cdot g\cdot {eV_{sd}}/{2} 
\label{eq:kTeff2}
\end{equation}
where $1\leq C(n)<2$.

As an application of the newly found distribution function, 
we investigate the current and the differential conductance 
 {\em G=dI/dV} of the SWNT-SET with $g=0.24$, as a function of gate voltage.
Current $I$ is 
\begin{equation}
\begin{array}{l}
\begin{displaystyle}
I = -e\hspace{-5mm} \sum_{N,\{n\},\{n^\prime\}}\hspace{-5mm}  \Big[P(N,\{n\})
\Gamma_L(N,\{n\} \rightarrow N+1,\{n^\prime\}) \end{displaystyle}\\
~~~  -P(N+1,\{n^\prime\})
\Gamma_L(N+1,\{n^\prime\}) \rightarrow N,\{n\})\Big].
\end{array}
\label{eq:I}
\end{equation}

\begin{figure}[htbp]
\begin{center}
\psfrag{xlab_T}{T[K]}
\psfrag{xylab_A10}{10}
\psfrag{xlab_A0}{$^{0}$}
\psfrag{xlab_A2}{$^{2}$}
\psfrag{xlab_BV}{V}
\psfrag{xlab_Bsd}{$_{sd}$}
\psfrag{xlab_BV2}{[V]}
\psfrag{ylab_A6}{$^{-6}$}
\psfrag{ylab_A7}{$^{-7}$}
\psfrag{ylab_AG}{G}
\includegraphics[width=6.5cm,height=9.0cm]{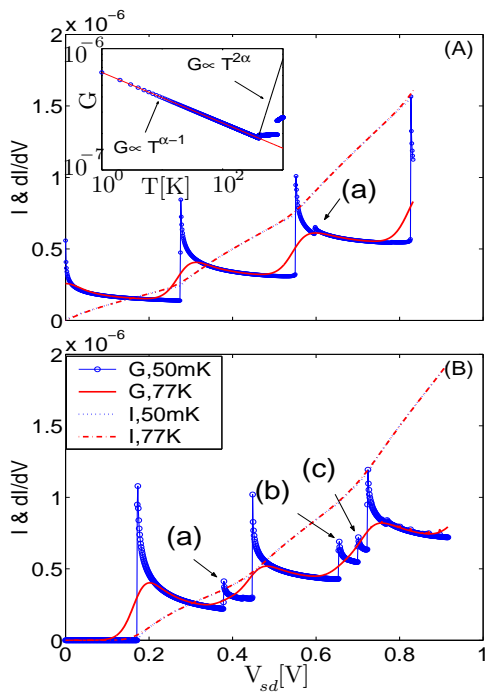}
\caption{Current $I$ and differential conductance $dI/dV$ as a function 
of source-drain voltage $V_{sd}$ for $g=0.24$. 
(A) Off-Coulomb blockade with  $N_G$=0.5 and (B) a Coulomb blockade
with { $N_G$=0.35}, for which the conductance vanishes up to 
$eV_{sd}=0.3\times\pi\hbar v_F/g^2L_D$ at zero temperature.
$I$ is in units proportional to $|t_{L/R}|^2$.
Inset in (A) shows power-law of conductance on temperature
at zero bias voltage where solid line is $G(T)=G(T_0)(T/T_0)^{\alpha-1}$ 
for $k_BT$$<$$\Delta E$.
Implicit parameters are same as in Fig. \ref{fig:PnallvsEn}.
}
\label{fig:IV}
\end{center}
\end{figure}

Fig. \ref{fig:IV} shows the {\em I-V} characteristics: (A) off-Coulomb blockade
with {$N_G$=0.5} and (B) a blockade case 
with { $N_G$=0.35} ($I$ is in units proportional to $|t_{L/R}|^2$). 

At zero temperature, there are conductance peaks whenever voltage drop across 
a junction equals difference between two quantized energy levels.
At a finite temperature only peaks that overcome thermal broadening survive. 
For {$N_G$=0.5} (Fig. \ref{fig:IV}(A)), electron numbers $N$=0 and $N$=1 have 
degenerate ground states which results in lifting of Coulomb blockade 
at zero bias voltage and yields equidistant conductance peaks separated by 
  $2\Delta E$=$2(\pi\hbar v_F/gL_D)$. 
In Fig. \ref{fig:IV}(B),
there is no transport at low bias voltage due to Coulomb blockade.
The main conductance peaks indicate transitions 
involving the ground state. The side peaks $(a)$ in Fig. \ref{fig:IV}(A) and
$(a)$ to $(c)$ in Fig. \ref{fig:IV}(B) are 
for transitions between excited states \cite{braggio01,jaeuk02b}.
The satellite peaks are smeared out at a high temperature as shown in the T=77K plot
because they result from transitions between excited states. 

The inset in Fig. \ref{fig:IV}(A) shows conductance at zero bias voltage is
proportional to a power of temperature, $G_{V\rightarrow 0^+}(T) \propto T^{\alpha^\star}$,
$\alpha^\star=\alpha-1$ when $k_BT$$<$$\Delta E$ and 
it is eventually expected to reach $ 2\alpha$ when $k_BT$$\gg$$ \Delta E$ where
the dot displays continuous spectrum.
It can be generalized to $G_{V \rightarrow V_{p}^+}(T) \propto T^{\alpha-1}$
when $k_BT$$<$$ \Delta E$.

The {\em G-V} analysis agrees with Braggio {\em et al.} \cite{braggio00,braggio01}, 
noticing that our system has four channel exponent,
$\alpha$=$(g^{-1}$$-1)/4$ and spin effects in the dot are not considered \cite{jaeuk03}. 
Since the results by Braggio {\em et al.} were obtained
in a different approximation for the occupation probabilities, 
this comparison also shows that a simple {\em I-V} measurement is rather insensitive 
to the on-dot distribution function.
However, the distribution function is 
expected to have a more pronounced impact on sensitive probes such as 
noise measurements, which will be considered at elsewhere.


In conclusion, we have analyzed a quantum wire single electron transistor
where each wire segment is described by a Tomonaga-Luttinger model. 
We have shown that in the sequential tunneling regime the
steady-state occupation probabilities of the many-body states
at the central dot follow a universal distribution that is determined
by a competition between energy dependent tunneling rates and the
degeneracy of the on-dot many-body quantum states. 
We have applied the non-equilibrium distribution to calculate the
current-voltage characteristics of a SWNT-SET, and shown that 
the {\em I-V} measurement is rather insensitive to the distribution function.

We acknowledge valuable discussions with R. I.  Shekhter, L. Y.  Gorelik
and A.  Kadigrobov, and the support of the Swedish Foundation for 
Strategic Research (QDNS and CARAMEL programs)
and the Royal Swedish Academy of Sciences. I. Krive acknowledges 
the hospitality of Dept. of Applied Physics, Chalmers University of
Technology and G\"oteborg University.


\bibliographystyle{plain}

\begin{thebibliography}{99}
\bibitem{jaeuk02b} Note that, in Fig. \ref{fig:IV}(B), the peak corresponding to
(N=0,n=j+1)$\leftrightarrow$(N=1,n=j) is missing at low $T$ as it results from 
the transitions between (nearly degenerate) excited states.
\bibitem{jaeuk02} A system with $g_{lead}$=$g$ and $g_{dot}$=1 shows that
the averaged probabilities follow the analytic result (\ref{eq:P_analytic})
and the probability fluctuations for various $g$-values
are almost identical to $g_D$=1 case (inset of Fig. \ref{fig:PnallvsEn}(A)). 
On the other hand,  a system with $g_{lead}$=$1$ and $g_{dot}$=$g$ shows fluctuation 
suppressions as a function of $g$, similar to the inset of Fig. \ref{fig:PnallvsEn}(B);
however, the averaged probabilities follow Eq. (\ref{eq:P_analytic}) with $\alpha$=0.
Finally, when the dot transition matrix elements in Eq. (\ref{eq:Laguerre}) 
are replaced by unity, it shows no fluctuation of the probabilities at all
and averaged probabilities are in a good agreement with Eq. (\ref{eq:P_analytic}) 
for the whole range of interaction strength.
\end{thebibliography}

\end{document}